\documentclass[11pt]{article}
    \usepackage[T1]{fontenc}
\def\beq{\begin{equation}}
\def\eeq{\end{equation}}
\usepackage{geometry}
\usepackage{amsfonts} 
\usepackage{amsmath}
\usepackage{amssymb}
\usepackage{graphicx}
\usepackage{url}
\usepackage{tikz}
\usetikzlibrary{matrix,arrows,shapes}
\usepackage{eucal}
\usepackage{dsfont}
\usepackage{bbm}
\usetikzlibrary{trees}
\usetikzlibrary{decorations.pathmorphing}
\usetikzlibrary{decorations.markings}
\definecolor{see}{RGB}{67,75,179}
\definecolor{darksee}{RGB}{42,44,148}
\definecolor{honey}{RGB}{232,226,129}
\definecolor{lighthoney}{RGB}{255,254,220}
\definecolor{dkbrown}{RGB}{70,47,36}
\definecolor{see}{RGB}{67,75,179}
\definecolor{darksee}{RGB}{42,44,148}
\definecolor{honey}{RGB}{232,226,129}
\definecolor{lighthoney}{RGB}{255,254,220}
\definecolor{dkbrown}{RGB}{70,47,36}
\definecolor{blue1}{RGB}{130,150,209}
\definecolor{blue2}{RGB}{42,60,122}
\definecolor{darkbgblue}{rgb}{0.01,0.05,0.15}
\definecolor{darkfgblue}{rgb}{0.25,0.45,0.5}
\definecolor{fgblue}{rgb}{0.2, 0.3, 0.5}
\definecolor{bgblue}{rgb}{0.92, 0.93, 0.95}
\newcommand{\E}{\mathfrak{E}}

\newcommand{\BV}{\mathfrak{BV}}
\newcommand{\fA}{\mathfrak{A}}

\newcommand{\D}{\mathfrak{D}}

\newcommand{\F}{\mathfrak{F}}

\newcommand{\Hcal}{\mathcal {H}}
\newcommand{\Lcal}{\mathcal {L}}

\newcommand{\Ncal}{\mathcal{N}}
\newcommand{\Mcal}{\mathcal{M}}
\newcommand{\Ocal}{\mathcal{O}}
\newcommand{\Scal}{\mathcal{S}}

\newcommand{\id}{\mathrm{id}}               
\DeclareMathOperator{\supp}{\mathrm{supp}}      
\newcommand{\loc}{\mathrm{loc}}
\newcommand{\inv}{\mathrm{inv}}

\newcommand{\RR}{\mathbb{R}}           
\newcommand{\M}{\mathbb{M}} 	     
\newcommand{\al}{\alpha}

\newcommand{\la}{\lambda}

\newcommand{\ph}{\varphi}

\newcommand{\sst}[1]{\scriptscriptstyle{#1}}  
\newcommand{\vr}[1]{\boldsymbol{#1}}         
\newcommand{\1}{\mathds{1}}                         
\newcommand{\be}{\begin{equation}}
\newcommand{\ee}{\end{equation}}

\newcommand{\Lap}{\bigtriangleup}

\let\affiliationfont\rhfont
\def\authorfont{\footnotesize}
\def\address#1{\par
	{\centering{\affiliationfont#1\par}}\par\vspace*{11pt}
}
\def\keywords#1{\par
	\vspace*{8pt}
	{\authorfont{\leftskip26pt\rightskip\leftskip
			\noindent{\it Keywords}\/:\ #1\par}}\vskip-12pt}

\begin{document}

\title{Effective quantum gravity observables\\ and locally covariant QFT}
\author{Kasia Rejzner$^*$}
\date{}
\maketitle
\address{Department of Mathematics, University of York,\\
York, YO10 5DD, England\\
$^*$E-mail: kasia.rejzner@york.ac.uk\\
http://rejzner.com/}
\begin{abstract}
Perturbative algebraic quantum field theory (pAQFT) is a mathematically rigorous framework that allows to construct models of quantum field theories on a general class of Lorentzian manifolds. Recently this idea has been applied also to perturbative quantum gravity, treated as an effective theory. The difficulty was to find the right notion of observables that would in an appropriate sense be diffeomorphism invariant. In this article I will outline a general framework that allows to quantize theories with local symmetries (this includes infinitesimal diffeomorphism transformations) with the use of the BV (Batalin-Vilkovisky) formalism. This approach has been successfully applied to effective quantum gravity in a recent paper by R. Brunetti, K. Fredenhagen and myself. In the same paper we also proved perturbative background independence of the quantized theory, which is going to be discussed in the present work as well. 
\end{abstract}




\section{Algebraic approach to QFT}
Quantizing gravity is one of the most challenging problems faced by modern theoretical physics. Among possible approaches, the most popular ones are the loop quantum gravity and the string theory. Despite the efforts of many decades the full theory of quantum gravity (QG) has not yet been established and the questions we face are of both technical and conceptual nature. Among the latter, one should mention the problem of identifying what should be the observables of quantum gravity. It turns out that contrary to earlier believes, this problem can be formulated and solved in the framework of quantum field theory on curved spacetimes. This observation has been made in Ref.~\cite{BFgrav} and was a motivation to take seriously the idea to quantize gravity as an effective field theory. The framework which allows to perform this task is that of perturbative algebraic quantum field theory (pAQFT). Let us start this article with an overview of pAQFT and its applications to building models of QFT's on curved backgrounds.

The algebraic approach to QFT goes back to the idea of Haag and Kastler Ref.~\cite{HK} (see also Ref.~\cite{Haag}) to formulate the axiomatic framework for theories of quantized fields, based on the concept of \textit{locality}. Originally AQFT was formulated as a theory on Minkowski spacetime $\M=(\RR,\eta)$, where $\eta=\mathrm{Diag}(1,-1,-1,-1)$ is the Minkowski metric. Later on it was generalized to a larger class of spacetimes $\Mcal=(M,g)$. 

We say that a curve $\gamma$ in a spacetime $\Mcal$ is timelike/null/spacelike if its tangent vector $\dot{\gamma}$ fulfills $g(\dot{\gamma},\dot{\gamma})>0$/ $g(\dot{\gamma},\dot{\gamma})=0$/ $g(\dot{\gamma},\dot{\gamma})<0$ respectively. A curve that is timelike or null is called causal. According to Einstein's general relativity, light moves on null curves and observers follow timelike curves, so there is no way to send information between spacelike separated regions. This principle seems to be in conflict with quantum mechanics, due to the existence of entanglement, but in fact, can be implemented in algebraic quantum field theory (AQFT).

\begin{figure}[b]
\begin{center}
	\begin{tikzpicture}[scale=0.37]
	\shadedraw[shading=radial, inner color=honey,outer color=lighthoney,scale=0.65] plot[smooth cycle] coordinates{(-1,6)(-3,7)(-4,9)(-3,10.7)(-1,11)(1,10)(2.5,9.8)(3.5,10)(4.9,10)(5.3,8.5)(3,6)(1,5.7)};
	\draw(0,5.5) node {$\fA(\Ocal_2)$};
	\draw (0.6,0.5) node[shading=axis,shading angle=90,  left color=blue1,right color=blue1!20!white,name=t,minimum size=2cm,draw,diamond] {};
	\draw(-1,0.5) node {$\Ocal_2$};
	\draw (1,0.5) node[name=s,minimum size=1cm,draw,diamond] {};
	\draw (7,0.5) node[shading=axis,shading angle=90,  left color=blue1,right color=blue1!20!white,name=t,minimum size=1cm,draw,diamond] {}
	node {$\Ocal_1$};
	\shadedraw[shading=radial, inner color=honey,outer color=lighthoney,scale=0.6] plot[smooth cycle] coordinates{(12,6)(10,7)(9,9)(10,10.7)(12,11)(14,10)(15.5,8)(14,6.2)};
	\path[->] (t.north east) edge  [bend right]  node[left] {} (7.5,4.5);
	\path[->] (-1.5,1) edge  [bend left]  node[left] {} (-2,5);
	\draw(7,5.5) node {$\fA(\Ocal_1)$};
	\draw(4.5,5.5) node  {$\supset$};
	\draw(4.5,0.5) node  {$\supset$};
	\end{tikzpicture}
\end{center}
\caption{Diagram illustrating the isotony axiom}\label{Fig1}
\end{figure}
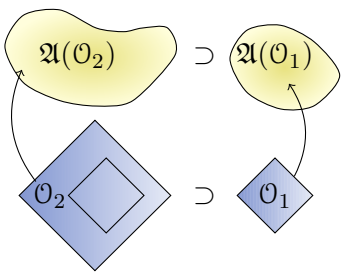

In the AQFT framework, a model is defined by specifying algebras of \textit{local} observables assigned to bounded regions $\Ocal$ of $\M$. The physical notion of subsystems is encoded in  the condition of \textit{isotony}. It means that if we have two bounded regions $\Ocal_1,\Ocal_2$ such that $\Ocal_1\subset\Ocal_2$, then $\fA(\Ocal_1)\subset\fA(\Ocal_2)$, i.e. we don't loose observables by going to a larger region  (see Fig. \ref{Fig1}.).
The net provides us only with algebras of observables of bounded regions, but one can construct algebras for more general regions by certain limiting procedures. For example, the algebra of the full spacetime is obtained as the inductive limit $\fA\doteq\overline{\bigcup\nolimits_\Ocal\fA(\Ocal)}$ (the bar means certain topological completion).

An assignment of observable algebras to bounded regions of $\M$ that fulfills the isotony requirement defines a \textit{net of algebras}. In the original framework these algabras were required to be $C^*$-algebras (the abstract generalization of the concept of the algebra of bounded operators on a Hilbert space), but this requirement has to be weakened if we want to use the perturbative methods to build models.

Other axioms required in the AQFT framework include:
\begin{itemize}
	\item {\bf Einstein causality}: intuitively, it implements the idea that ``nothing travels faster than light'', so the measurements performed in spacelike regions should be independent. More precisely, if $\Ocal_1$ is spacelike to $\Ocal_2$ (i.e. there is no causal curve connecting any point of $\Ocal_1$ with a point of $\Ocal_2$), then $[A,B]=0$ for any $A\in\fA(\Ocal_1)$, $B\in\fA(\Ocal_2)$ and the commutator is taken in the sense of $\fA(\Ocal_3)$, where $\Ocal_3$ is any bounded region that contains both $\Ocal_1$ and $\Ocal_2$.
	\item {\bf Time-slice axiom} is the quantum version of the well posedness of the Cauchy problem in classical theory. More precisely, the algebra of observables localized in a thin time-slice is isomorphic to the full algebra $\fA$. 
\end{itemize}

There are further axioms that capture some further physical features of the theory, but in our context the two listed above are most relevant, since they allow for a straightforward generalization to curved spacetimes. Before we make this step, let us discuss the problem of physical interpretation of the theory and how it relates to the notion of a \textit{state}. In quantum mechanics one usually starts with a Hilbert space and then defines observables as operators acting on this space. In the AQFT framework one takes a more abstract viewpoint by starting with algebras of observables. The physical interpretation is obtained by specifying \textit{a state}, which physically corresponds to the way we prepare the experimental setup. Mathematically, a state on a  $*$-algebra $\fA$ (algebra with the involution operation $*$) is a linear functional $\omega$, such that:
\[ \omega(A^*A) \geq 0,\  \omega(\1) = 1\,.\]
Given a $*$-algebra one can associate to a state $\omega$ on it a Hilbert space representation $\Hcal_\omega$, using the GNS theorem (for a pedagogical introduction into algebraic approach to quantum theory see for example Refs.~\cite{Moretti,LesHouches}). This fact provides a link between the AQFT setting and the more commonly used Hilbert space language.

The crucial feature of AQFT is that the observables are local (obey the Einstein causality axiom), but the states are not, so they contain the information about possible correlations. In fact, it has been shown in Ref.~\cite{WernerSummers} that in some simple examples of QFT models, the vacuum state is maximally entangled. By separating the concept of states from the concept of observables, AQFT allows to deal with the apparent contradiction between causality and entanglement. This feature is also very useful in the context of QFT on curved spacetime.
\section{Locally covariant quantum field theory}\label{LCQFT}
Let us briefly recall the difficulties one has to face when constructing QFT models on curved backgrounds. Firstly, the group of spacetime symmetries of a generic spacetime is trivial. It follows that the very concept of particles in the sense of Wigner is no longer available and the idea of the vacuum as the state with no particles becomes meaningless. Another difficulty arises from the fact that transition to imaginary times (and a corresponding transition to a Riemannian space) is possible only in special cases. Moreover, the Fourier transform is in general not defined, so calculations usually performed in momentum space cannot be done.  All these facts lead to  some peculiarities of quantum field theory on generic spacetimes, which include: particle creation, Hawking radiation and the Unruh effect.

The conceptual difficulty related to the lack of the distinguished vacuum state is resolved in the AQFT framework by the fact that the algebra of observables can be defined abstractly (no need to start with any distinguished state) using only the local data. In the next step one looks for some physically motivated states on this algebra. Therefore, one can replace $\M$ with a more general Lorentzian manifold $\Mcal=(M,g)$ and define a QFT model by constructing a net of local algebras satisfying the Einstein causality and the time-slice axiom. For the latter, and for some other technical reasons, it is necessary to require $\Mcal$ to be \textit{globally hyperbolic} (contains a Cauchy surface).

We can go a step further and see what happens if we replace the embeddings of bounded regions $\Ocal$ into a fixed spacetime $\Mcal$ with arbitrary embeddings between pairs of globally hyperbolic spacetimes $\Ncal$ and $\Mcal$. We formalize this idea by introducing the notion of an \textit{admissible embedding}. An embedding $\chi:\Mcal\to \Ncal$ of a globally hyperbolic manifold $\Mcal$ into another one $\Ncal$ is \textit{admissible} if it is an isometry and it preserves orientations and the causal structure. The property of \textit{preserving the causal structure} is defined as follows: let $\chi:\Mcal\to \Ncal$, for any causal curve $\gamma : [a,b]\to \Ncal$, if $\gamma(a),\gamma(b)\in\chi(\Mcal)$ then for all $t	\in ]a,b[$ we have: $\gamma(t)\in\chi(\Mcal)$.

As in the original AQFT framework, we assign algebras of observables to globally hyperbolic spacetimes and we also want to require that for each such admissible embedding there exists an injective homomorphism 
\be
\alpha_{\chi}:\mathfrak{A}(\Mcal)\to\mathfrak{A}(\Ncal)
\ee
of the corresponding algebras of observables assigned to them, moreover if $\chi_1:\Mcal\to \Ncal$ and $\chi_2:\Ncal\to \Lcal$ are embeddings as above, then we require the covariance relation
\be\label{cov}
\alpha_{\chi_2\circ\chi_1}=\alpha_{\chi_2}\circ\alpha_{\chi_1} \ .
\ee
The two axioms mentioned in the previous section in the context of AQFT are easily generalized to the LCQFT setting.
\begin{itemize}
	\item \textbf{Einstein causality}: let $\chi_i:\Mcal_i\rightarrow \Mcal$, $i=1,2$ be admissible embeddings such that $\chi_1(M_1)$ is spacelike separated from $\chi_2(M_2)$, then we require that:
	\[
	[\al_{\chi_1}(\fA(\Mcal_1)),\al_{\chi_2}(\fA(\Mcal_2))]=\{0\}\,,
	\]
	\item \textbf{Time-slice axiom}: let $\chi:\Ncal\rightarrow \Mcal$ be an admissible embedding, if $\chi(\Ncal)$ contains a neighborhood of a Cauchy surface $\Sigma\subset \Mcal$, then $\al_\chi$ is an isomorphism.
\end{itemize}

The next important notion is that of a \text{quantum field}. In the LCQFT framework the role of fields is to provide labels for observables. More precisely, we want to compare observables measured in different regions of a spacetime $\Mcal$ and this cannot be done in a simple way if $\Mcal$ has no non-trivial isometries. In order to make our QFT model useful, we need to be able to say what physical quantities (e.g. temperature, local energy density) are represented by given observables.

\begin{figure}[t]
 \begin{center}
	\begin{tikzpicture}[scale=0.65]
	\shadedraw[shading=axis,shading angle=90, left color=blue1,right color=white] plot[smooth cycle] coordinates{(-1,-1) (-2.2,0)(-2.1,1)(-1,2)(1,2.3)(2,2.5)(3,2)(3,1)(2,0)(1,-1)(0,-1.2)};
	\draw(-1.4,0.5) node{$\Mcal$};
	\draw(4.3,2) node [name=F]{$\Phi_\Mcal(\psi_*f)$};
	\draw(4.3,-3.5) node [name=F2]{$\Phi_{\Ocal}(f)$};
	\draw (.7,0.5) node[name=s,minimum size=2cm,draw,diamond, inner sep=0pt] {};
	\draw (.7,-5) node[shading=axis,shading angle=90, left color=blue1,right color=white,name=t,minimum size=2cm,draw,diamond] {};
	\draw(.7,0.3) node {\small{$\psi(\Ocal)$}};
	\draw(.7,-6) node {\small{$\Ocal$}};
	\shade[shading=radial, inner color=darksee,outer color=blue1,scale=0.24] plot[smooth cycle] coordinates{(3.5,-21.4)(1.5,-20.4)(0.5,-19.4)(1.5,-17.7)(3.5,-17.4)(5.5,-18.4)(6.4,-20.2)(5.5,-21.2)};
	\shade[shading=radial, inner color=darksee,outer color=blue1,scale=0.24] plot[smooth cycle] coordinates{(3.5,2.6)(1.5,3.6)(0.5,4.6)(1.5,6.3)(3.5,6.6)(5.5,5.6)(6.4,3.8)(5.5,2.8)};
	\draw(.7,1.2) node [name=f,color=white]{\small{$\psi_*f$}};
	\draw(.7,-4.5) node [name=f2,color=white]{\small{$f$}};
	\path[->] (t.north east) edge  [bend right,thick]  node[right] {$\psi$} (s.south east);
	\path[<-] (t.north west) edge  [bend left]  node[left] {$\psi^{-1}$} (s.south west);
	\path[->] (f) edge  [color=red, bend left]  node[right] {} (F);
	\path[->] (f2) edge  [color=red, bend right]  node[right] {} (F2);
	\end{tikzpicture}
		\end{center}
		\caption{Diagram illustrating the covariance condition for locally covariant fields.}\label{Fig2}
		\end{figure}
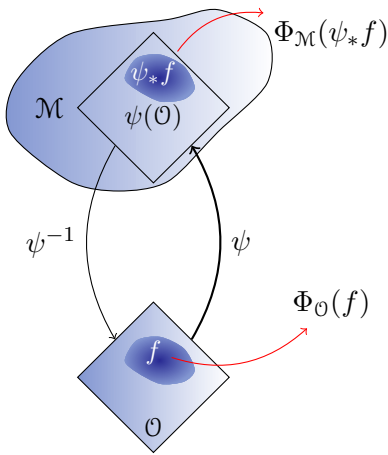

To make this more precise mathematically, let us denote by $\D(\Mcal)$ the space of test functions on $\Mcal$. A \textit{locally covariant field} is a family of maps $\Phi_\Mcal:\D(\Mcal)\rightarrow \fA(\Mcal)$, labeled by spacetimes $\Mcal$ such that: 
\[
\alpha_\psi(\Phi_\Ocal(f))=\Phi_{\Mcal}(\psi_*f)\,,
\]
where $\psi:\Ocal\rightarrow\Mcal$ is an admissible embedding  (see Fig. \ref{Fig2}.). The notion of locally covariant fields generalizes the notion of Wightman's operator-valued distributions (note the dependence on test functions).
For more detail about locally covariant QFT, see for example Refs.~\cite{FewVerchReview,FR14}.
\section{Effective quantum gravity}
The road to effective quantum gravity from locally covariant quantum field theory is not easy.
It is paved with numerous technical and conceptual problems and it took a few decades before the suitable mathematical tools became available.
\subsection{Outline of the approach} 
  In contrast to QFT on curved spacetimes, in quantum gravity (QG) the spacetime structure is dynamical. This means that we cannot treat the metric as a fixed structure, but it interacts with the matter field. One can partially model this situation using the framework involving \textit{backreaction}. In this formalism one treats matter fields as quantum objects and studies their effect on the metric by inserting the expectation value of the quantum stress-energy tensor in a given state $\omega$ into Einstein's equations:
\[
\left<T_{\mu\nu}\right>_{\omega}=G_{\mu\nu}\,,
\]
where $G_{\mu\nu}=R_{\mu\nu}-\frac{1}{2}Rg_{\mu\nu}$ is the Einstein tensor. In the pAQFT framework this approach has been applied in cosmology and in the study of QFT in black-hole spacetimes, see for example Refs.~\cite{Hack14,DMP09,DMP11} and a recent book, Ref.~\cite{ThomasBook}.

On the next level of approximation one splits the metric $g$ into the background metric $g_0$ and a perturbation $h$ and quantizes the perturbation as a quantum field on the background $g_0$. This is the approach which was taken in Ref.~\cite{BFRej}. Since this tentative split into background and perturbation is not physical, one needs to show that the predictions of the theory do not depend on the way $g$ is split. This consistency condition is called \textit{background independence}. In the pAQFT approach, the background independence of effective QG was proven in Ref.~\cite{BFRej} in the sense that a localized change in the background which yields an automorphism on the algebra of observables (called relative Cauchy evolution in Ref.~\cite{BFV}) is actually trivial, in agreement with the proposal made in Ref.~\cite{BFgrav} (see also Ref.~\cite{FRindep}). We will come back to this issue in subsection \ref{Bindep}.

Another conceptual difficulty in quantizing gravity is that the Einstein-Hilbert action is reparametrization invariant, hence the theory has a huge symmetry group, the diffeomorphism group. This means that labeling of spacetime points doesn't have a physical meaning. As a consequence, physical observables have to be diffeomorphism invariant. In the framework of  
Ref.~\cite{BFRej} the characterization of diffeomorphism invariant observables is given by means of the BV formalism.

Finally, there is a known difficulty that quantum gravity, as a QFT, is power counting non-renormalizable. We deal with this problem by using the Epstein-Glaser renormalization scheme, which allows us to calculate finite contributions to  renormalized time-ordered products to every order in $\hbar$ and the coupling constant. The theory is then interpreted as an effective theory with the property that only finitely many parameters have to be considered below a fixed energy scale (see Ref.~\cite{Weinberg}). Another possible direction would be to make contact with the \textit{asymptotic safety} approach. A theory is called asymptotically safe if there exists an ultraviolet fixed point of the renormalisation group flow with only finitely many relevant directions (see Ref.~\cite{Weinberg79}). Results supporting this perspective have been obtained by Reuter and Saueressig in Refs.~\cite{Reu98,Reu02}.
\subsection{Building models in pAQFT}
In this article we focus on the notion of observables, following the ideas introduced in Ref.~\cite{BFRej}. The framework used in this work is that of perturbative algebraic quantum field theory (pAQFT). In this framework the axioms introduced in the previous two sections are still valid, but one works with algebras that are formal power series in $\hbar$ and the coupling constant $\la$. The construction of pAQFT models can be summarized as follows:
\begin{enumerate}
	\item Construct the classical theory using the Lagrangian $L$.
	\item Split the Lagrangian into the free part $L_0$ and the interacting part $L_I$. Quantize $L_0$ using deformation quantization.
	\item Introduce the interaction using the Epstein-Glaser renormalization scheme. 
\end{enumerate}
If local symmetries are present, as is the case in Yang-Mills theories and general relativity, the starting Lagrangian $L$ has to be extended with some auxiliary fields like ghosts, antighosts and antifields. The systematic way to do this is provided by the BV (Batalin-Vilkovisky) formalism. The precise formulation of the BV framework adapted to pAQFT has been given in Refs.~\cite{FR,FR2}.

The implementation of the classical Lagrangian formalism which we use in pAQFT differs slightly from the commonly used one, so we will briefly review its main features. Firstly, the Lagrangian for us is a locally covariant classical field, as defined in section \ref{LCQFT}, valued in off-shell functionals on the configuration space. Let us clarify this in more detail. Let $\E(\Mcal)$ be the off-shell configuration space of the theory. For effective quantum gravity we take $\E(\Mcal)=\Gamma((T^*M)^{\otimes_s 2})$, the space of covariant symmetric 2-tensors. On this space we consider the space $\F_{\loc}(\Mcal)$ of smooth functionals that are local, i.e. depend on the field configuration at a given point only via the finite jet. More precisely, they are of the form
\[
F(h)=\int_M \omega(j_x^k(h))\,,
\]
where $j_x^k(h)$ is the $k$-the jet prolongation of $h$ and $\omega$ is a density-valued function on the jet bundle.

Sums of products of local functionals are called multilocal and we denote the space of multilocal functionals by $\F(\Mcal)$. An important property characterizing a functional $F\in\F(\Mcal)$ is its spacetime support defined by
\begin{align}\label{support}
\supp F\doteq\{ & x\in M|\forall \text{ neighborhoods }U\text{ of }x\ \exists h_1,h_2\in\E(\Mcal), \supp h_2\subset U\,,
\\ & \text{ such that }F(h_1+h_2)\not= F(h_1)\}\ .\nonumber
\end{align}
We will assume that all our functionals are compactly supported. 

We are now ready to introduce generalized Lagrangians.
In our framework a generalized Lagrangian is a family of maps $L_{\Mcal}:\D(\Mcal)\rightarrow \F_{\loc}(\Mcal)$ satisfying the covariance condition:
\[
L_\Ocal(f)[\chi^*h]=L_{\Mcal}(\psi_*f)[h]\,,
\]
where $\psi:\Ocal\rightarrow\Mcal$ is an admissible embedding, $h\in\E(\Mcal)$, $f\in\D(\Ocal)$. Moreover, we require
\be\label{L:supp}
\supp(L_\Mcal(f))\subseteq \supp(f)\,,\quad \forall f\in\D(\Mcal)\,,
\ee
and the additivity rule 
\be\label{L:add}
L_\Mcal(f_1+f_2+f_3)=L_\Mcal(f_1+f_2)-L_\Mcal(f_2)+L_\Mcal(f_2+f_3)\,,
\ee
for $f_1,f_2,f_3\in\D(\Mcal)$ and $\supp f_1\cap\supp f_3=\emptyset$.
Intuitively, we think of a generalized Lagrangian as a Lagrangian density smeared with a cutoff function. In the traditional Lagrangian classical field theory we would replace the smooth cutoff function with a characteristic function of a region situated between two Cauchy surfaces. This, however, would lead to some unpleasant divergences in quantization, so we prefer the smoothed-out version.

The action $S(L)$ is defined as an equivalence class of Lagrangians  (see Ref.~\cite{BDF}), where two Lagrangians $L_1,L_2$ are called equivalent $L_1\sim L_2$  if
\be\label{equ}
\supp (L_{1,\Mcal}-L_{2,\Mcal})(f)\subset\supp\, df\,, 
\ee
for all spacetimes $\Mcal$ and all $f\in\D(\Mcal)$. In general relativity the dynamics is given by the Einstein-Hilbert Lagrangian:
\be\label{EH}
L^{\sst EH}_{(M,g_0)}(f)(h)\doteq \int R[g]f\,d\mu_{ g},\quad h\in\E(\Mcal)\,,\ g=g_0+h\,,
\ee
where we use the Planck units, so in particular the gravitational constant $G$ is set to 1.

The equations of motion are given in terms of the Euler-Lagrange derivative of $S(L)$ defined by
\[
\left<S'_\Mcal(h_0),h\right>\doteq \left<L^{(1)}_\Mcal(f)[h_0],h\right>\,,
\]
where $h_0\in\E(\Mcal)$, $h\in\E_c(\Mcal)$ is a compactly supported configuration, and $f\in\D(\Mcal)$ is chosen such that $f\equiv 1$ on $\supp(h)$. Since $L$ is local, the definition of $S'$ is independent of the choice of the cutoff function with the above property and we define the equation of motion as
\[
S'(h_0)=0\,.
\]

The classical theory can be defined by introducing the Poisson bracket using the covariant Peierls method, as proposed in Ref.~\cite{Pei}. This method relies on the existence of unique retarded and advanced Green's functions for the linearized equations of motion. This would not be possible for theories with local symmetries (like gravity) without some further steps. 

As mentioned at the beginning of this section, we use the BV formalism to introduce the auxiliary degrees of freedom: ghosts $c\in\Gamma(TM)[1]$, antighosts $\overline{c}\in\Gamma(TM)[-1]$, Nakanishi-Lautrup fields $b\in\Gamma(TM)[0]$ and antifields for all degrees of freedom. The numbers in square brackets indicate the grading. The resulting extended configuration space is a graded manifold
\[
\overline{\E}(\Mcal)=\E(\Mcal)\oplus\Gamma(TM)[-1]\oplus\Gamma(TM)[0]\oplus\Gamma(TM)[1]\,.
\]
Next, one takes the odd cotangent bundle $T^*[-1] \overline{\E}(\Mcal)$ of this manifold and considers the space of multilocal functionals on it. This space, denoted by $\BV(\Mcal)$, is the underlying algebra of the BV complex. The fiber of $T^*[-1] \overline{\E}(\Mcal)$ can be parametrized using abstract generators called \textit{antifields} and denoted by $h^\ddagger$, $c^\ddagger$, $\overline{c}^\ddagger$ and $b^\ddagger$, so there is one antifield for each field in the theory. Note that functionals on $T^*[-1] \overline{\E}(\Mcal)$ can be thought of simply as multivector fields and therefore $\BV(\Mcal)$ is equipped with a natural bracket, namely (minus) the Schouten bracket $\{.,.\}$, defined by
\begin{itemize}
	\item $\{X,F\}=-\partial_XF$ for $X$ a vector field and $F$ a function,
	\item $\{X,Y\}=-[X,Y]$ for two vector fields $X,Y$,
	\item $\{.,.\}$ fulfills the graded Leibniz rule.
\end{itemize}

In the next step one introduces a differential $s$ on $\BV(\Mcal)$ (the classical BV differential), which contains the information about the gauge symmetries and the equations of motion. To this end we add some extra terms to the original Lagrangian $L^{\sst EH}_{\Mcal}$ and obtain the extended Lagrangian $L$, which now depends on both fields and antifields. The classical BV differential is defined by
\[
sX=\{X,L(f)\}\,,
\]
where $f\equiv 1$ on the support of $X$, so the differential is locally generated by the extended Lagrangian. The construction is done in such a way that the cohomology of the complex $(\BV(\Mcal),s)$ contains information about the gauge orbits and the solution space so that $H^0(\BV(\Mcal),s)$ is the space of gauge invariant on-shell functionals $\F^{\inv}_S(\Mcal)$.

The advantage of using the BV formalism is that $(\BV(\Mcal),s)$ is relatively easy to quantize using the steps given at the beginning of this section. For details see Ref.~\cite{FR2}. Quantum observables are recovered as the cohomology of the quantum BV operator $\hat{s}$ which is obtained from $s$ by a certain deformation (see subsection \ref{EGren} for more detail). Abstractly, this already provides the characterization of observables in any theory that fits into the pAQFT framework, including general relativity. The caveat is that we need to treat it as an effective theory, since we do not control the convergence in  the coupling constant. A more dire question is if the space of observables characterized abstractly as a certain cohomology is non-empty. This can be resolved only if we are able to provide concrete examples of observables that meet the requirements. An explicit construction has been proposed in Ref.~\cite{BFRej} and here we will review the main ideas.
\subsection{Gauge-invariant observables}
First we note that locally covariant fields introduced in section \ref{LCQFT} can be \textit{diffeomorphism equivariant}. Given an infinitesimal diffeomorphism $\xi\in\Gamma(TM)$ its action $\rho$ on a locally covariant field $\Phi$ evaluated on the spacetime $\Mcal$ is given by
\[
(\rho(\xi)\Phi_{\Mcal})(f)[h]=\Phi_{\Mcal}(\pounds_\xi f)[h]+\Phi_{\Mcal}(f)[\pounds_\xi h]\,.
\]
The condition of diffeomorphism equivariance is formulated as the requirement that
\[
\rho(\xi)\Phi_{\Mcal}\equiv 0\,,
\]
for all $\xi\in\Mcal$ and for all globally hyperbolic spacetimes $\Mcal$. All fields that are locally and covariantly constructed from the full metric $g=g_0+h$ are equivariant in the above sense (recall that $g_0$ is the background and $h$ the perturbation). For example the Einstein-Hilbert Lagrangian \eqref{EH} is such a filed.

To pass from equivariant fields to gauge-invariant observables we need to make the test function depend on the physical fields. This is related to the fact that, physically, points of spacetime have no meaning. To realize this in our formalism we have to allow for a freedom of changing the labeling of the points of spacetime. For simplicity, we restrict here the class of spacetimes we consider to spacetimes which admit a global coordinate system. 
We realize the choice of a coordinate system by introducing four scalar fields $X^\mu$, which parametrize points of spacetime. We can write any test function  $f\in\D(\Mcal)$ in the coordinate basis induced by $X$. Conversely, if we fix $\vr{f}\in\RR^4\rightarrow\RR$, then the change of $f=X^*\vr{f}$ due to the change of the coordinate system is realized through the change of scalar fields $X^\mu$. For a locally covariant field $\Phi$ we obtain a map
\[
\Phi_{\Mcal\vr{f}}(g,X)\doteq \Phi_\Mcal(X^*\vr{f})(g)\,,
\]
As long as $\Mcal$ is fixed, we will drop the subscript $\Mcal$ in $\Phi_{\Mcal\vr{f}}$ and use the notation $\Phi_{\vr{f}}$ instead. For example,
the Einstein-Hilbert action induces a map
\[
L^{\sst EH}_{\vr{f}}(g,X)=\int_M R[g](x)\vr{f}(X(x))d\mu_{g}(x).
\]

In order to get the correct transformation under diffeomorphisms, we have to replace $X^\mu$ with some scalars $X_g^\mu$, $\mu=0,\ldots,3$, which depend locally on the metric or matter fields, if the latter are present in the model. The particular choice of these fields is not relevant for the present discussion. They could be, in pure gravity, scalars constructed from the Riemann curvature tensor and its covariant derivatives (see  Refs.~\cite{Berg,BergKom}). However some particularly symmetric spacetimes do not admit such metric dependent coordinates, since in such cases the curvature might vanish (for a detailed discussion see Refs.~\cite{CHP09,CH10}), but this is a non-generic case. Moreover, if we consider pure gravity without matter fields, such highly symmetric spacetimes are physically not observable, because there is no way to probe their curvature operationally. If matter fields are present, one can construct $X^\mu$'s using them. A known example is the Brown-Kucha\v{r} model (Ref.~\cite{BrK}), which uses dust fields. Here, following Ref.~\cite{BFRej}, we briefly discuss a similar Ansatz, where the gravitational field is coupled to 4 scalar massless fields. We add to the Einstein-Hilbert action a term of the form
\[
L^{\sst KG}(f)(g,\phi^0,\dots,\phi^3)=\sum_{\al=0}^{3}\int_M (\nabla_g \phi^\al)^2 d\mu_{g}.
\]
The additional scalar fields satisfy the equations of motion
\[
\Box_g \phi^\al=0,\ \al=0,\dots,3\,.
\]
Classically, we can now identify the coordinate fields with the matter fields $\phi^\al$, i.e. we set $X_{g,\phi}^\mu=\phi^\mu$, $\mu=0,\dots,3$. With quantization in mind, we make the split of $g$ and $\phi^\al$ into background and perturbations, which will subsequently be treated as quantum fields. We set $g=g_0+\la h$ and $\phi^\al=\ph_0^\al+\la \ph^\al$. We write coupling constant $\lambda$ explicitly to make it easier to keep track of orders of the power series. Our gauge-invariant observables are of the form
\[
\Phi_{\vr{f}}(h,\ph^0,\dots,\ph^3)=\Phi_{(M,g_0)}(\phi^*\vr{f})(\la h)\,,
\]
where $\phi^*\vr{f}(x)\doteq \vr{f}(\phi^0(x),\dots,\phi^3(x))$. As a concrete example consider
\[
\Phi_{\vr{f}}(h,\ph^0,\dots,\ph^3)=\int_M R_{\mu\nu\al\beta}R^{\mu\nu\al\beta}[g_0+\la h]\vr{f}((\ph_0^0+\la\ph^0)(x),\dots,(\ph_0^3+\la\ph^3)(x))d\mu_{g_0+\la h}\,,
\]
where $\ph_0^\al$ define harmonic coordinates with respect to the background metric, i.e. $\Box_{g_0}\ph_0^\al=0$, $\al=0,\dots 3$ and we choose $\vr{f}$ such that $\ph_0^*\vr{f}$ is compactly supported. Note that the support of $\Phi_{\vr{f}}$ is equal to the support of $\ph_0^* f$, so is compact, as required in our formalism. 
The physical interpretation of the scalar fields $\phi^\al$ has to be made clear in concrete examples. A possible direction is applying this framework to cosmology.

The notion of observables we propose  captures the relations between given fields of the theory rather than absolute values of these fields at some spacetime points. This motivates the name
\textit{relational observables}. Note that they are conceptually similar to the notion of observables  introduced by Rovelli in the framework of loop quantum gravity (Ref.~\cite{Rovelli:2001bz}) and later used and further developed  in Refs.~\cite{Dittrich:2005kc,Thiemann:2004wk}.

Going a step furhter, there is no need to distinguish between the curvature invariants that enter the definition of $X_g$'s and those which are used to construct the density $\Phi_x$ in  $\Phi^\beta_{\vr{f}}(g)=\int_M\Phi_x(g)\vr{f}(X_g(x))$. Instead, one can consider a family of $N$ scalar  curvature invariants $R_1,\ldots, R_N$ and a class of globally hyperbolic spacetimes characterized by the 4-dimensional images under this $N$-tuple of maps. It was proven in Ref.~\cite{MuSa} that every globally hyperbolic spacetime with a time function $\tau$ such that $|\nabla\tau|\ge1$, can be isometrically embedded into the $N$-dimensional Minkowski spacetime $\M^N$ for a sufficiently large $N$ (fixed by the spacetime dimension). It follows that, depending on the physical model at hand, one can choose $N$ and construct  $R_1,\ldots, R_N$  in such a way that all spacetimes of interest are characterized uniquely in this setup. One can then consider observables of the form
\[
\int_{M}\vr{f}(R_1(x),\ldots, R_N(x))\,,
\]
where $\vr{f}:\M^N\rightarrow\Omega^4(M)$ is a density-valued function, assumed to be compactly supported inside the image of $M$ under the embedding $\ph:M\rightarrow \M^N$ defined by the family $R_1,\ldots,R_N$. One could then quantize the metric perturbation, in the same way as it was done in Ref.~\cite{BFRej} or quantize the embedding $\ph$ itself, as it was done for the bosonic string quantization in Ref.~\cite{BRZ}.
\subsection{The role of deformation quantization}
As in any pAQFT model, quantization of the effective theory of gravity starts from quantization of its linearized version. We use the split  of the metric $g$ into background $g_0$ and perturbation $h$ to expand the BV-extended Lagrangian $L$ into a Taylor series. The constant term can be neglected and if $g_0$ is a solution to Einstein's equations, then the linear term vanishes and the lowest non-trivial contribution is quandratic in $h$. We take the antifield number zero term of this quadratic contribution and call it \textit{the linearized free Lagrangian} denoted by $L_0$. We define the interction term as $L_I=L-L_0$. Clearly, the split of $L$ into $L_0$ and $L_I$ depends on the choice of $g_0$. However, in section $\ref{Bindep}$ we show that physical quantities do not depend on this split (background independence).

The free theory corresponding to $L_0$ can be quantized by means of deformation quantization. We introduce the star product $\star$ using a Moyal type formula (see Refs.~\cite{BDF,DF}). There are some technical subtleties related to the domain of definition and uniqueness of $\star$, but these are essentially the same in effective quantum gravity as in the scalar field theory and have been dealt with in Ref.~\cite{BDF}. Having defined the free theory, we need to ``put back'' the interaction. This is explained in the next subsection.
\subsection{Few words about Epstein-Glaser renormalization}\label{EGren}
The main technical ingredient we need in order to define interacting quantum fields in effective QG is Epstein-Glaser renormalization. Its conceptual basis differs from other commonly used renormalization schemes, since it doesn't require one to manipulate ill-defined divergent quantities. Instead, we work on the level of S-matrices and postulate axioms that an S-matrix of the given theory has to obey. These are the Epstein-Glaser axioms, summarized as follows (see Ref.~\cite{BDF}):
\begin{enumerate}
	\item {\bf Causal factorization}: $\Scal(F+G)=\Scal(F)\star\Scal(G)$, if $\supp(F)$ is later than $\supp(G)$.
	\item {\bf Starting element}: $\Scal(0)=1$, $\Scal^{(1)}(0)=\id$.
	\item {\bf Field independence}: $\frac{\delta}{\delta \ph}\Scal(F)=\Scal^{(1)}(F)[\frac{\delta}{\delta \ph} F]$, where $\ph$ can be $h$, $c$, $\overline{c}$ or $b$, i.e. $\Scal$ doesn't explicitly depend on the field configurations.
	\item {\bf Unitarity}: $\overline{\Scal(\overline{F})}\star \Scal(F)=1$.
\end{enumerate}
There are some further conditions, but we do not state them here to keep the discussion as non-technical as possible. The main theorem of Epstein and Glaser (Ref.~\cite{EG}) shows the existence of S-matrices fulfilling the above axioms for local $F$ in scalar field theory on $\M$. This result has been generalized to curved spacetimes in Refs.~\cite{BF0,HW,HW01}. The non-uniqueness of the S-matrix for a given theory is related to the renormalization freedom. The relation of the resulting renormalization group to the Wilsonian renormalization group and the Polchinski flow equation has been shown in Ref.~\cite{BDF}. 

The generalization to gauge theories requires some additional renormalization conditions called Ward identities. These were proven for Yang-Mills theories on curved spacetimnes by Hollands in Ref.~\cite{H} and generalized in Ref.~\cite{FR2} to a larger class of theories with local symmetries that includes effective gravity. It was also shown in  Ref.~\cite{FR2} that some of the Ward identities can be summarized in the requirement that
\[
\{\Scal(L_I(f)),L_0(f_0)\}=0
\]
in the algebraic adiabatic limit if $f_0\equiv 1$ on the support of $f$. Morally, the algebraic adiabatic limit means that we work with equivalence classes of Lagrangians modulo the equivalence relation \eqref{equ}. The above condition on the S-matrix can be rewritten as the \textit{quantum master equation}:
\[
\frac{1}{2}\{L_0(f_0)+L_I(f),L_0(f_0)+L_I(f)\}-i\hbar \Lap(L_I(f))\sim 0\,,
\] 
where $\Lap(L_I(f))$ is a local functional corresponding to the anomaly.  If the anomaly can be removed using the remaining renormalization freedom and $\{L_0(f_0)+L_I(f),L_0(f_0)+L_I(f)\}\sim 0$, then the quantum master equation is fulfilled. In Ref.~\cite{H} it has been shown that the procedure of anomaly removal works in Yang-Mills theories and in Ref.~\cite{BFRej} this result was generalized to effective quantum gravity.

Interacting quantum fields are defined by means of the Bogoliubov formula. For a bounded region $\Ocal\subset\Mcal$ we choose a test function $f\in\D(\Mcal)$ such that $f\equiv 1$ on $\Ocal$ and $F\in\F(\Mcal)$ is supported inside $\Ocal$. We define the interacting field corresponding to $F$ by
\be\label{Bogform}
F_{\textrm{int}}=-i\hbar\left.\frac{d}{dt}\left(\Scal(L_I(f))^{-1}\star\Scal(L_I(f)+tF)\right)\right|_{t=0}\,,
\ee 
where the inverse of $\Scal$ is the $\star$-inverse. The algebra generated by such interacting fields with respect to $\star$ is independent of the choice of $f$, up to an isomorphism. We call it $\fA_{\mathrm{int}}(\Ocal)$, the local algebra of interacting quantum fields localized in $\Ocal$. Assigning algebras $\fA_{\mathrm{int}}(\Ocal)$ to all the bounded regions $\Ocal\subset\Mcal$ defines the interacting net.

The quantum BV operator $\hat{s}$ is defined in such a way that
\[
s(F_{\textrm{int}})=(\hat{s}F)_{\textrm{int}}\,.
\]
The nilpotency of this operator is guaranteed by the quantum master equation, so its cohomology is well defined and it characterizes the space of gauge invariant quantum observables.
\subsection{Background independence}\label{Bindep}
The last thing to check is the background independence. During the construction of the effective quantum gravity model with pAQFT methods, we make a split of the metric $g$ into the background $g_0$ and perturbation $h$ and use this split to expand the full interacting Lagrangian as $L=L_0+L_I$ by means of the Taylor expansion. Now we want to see what will happen if we slightly perturb the background. If the theory is background independent, then physical quantities do not change under such a perturbation. Following Ref.~\cite{BFRej} we sketch the argument that this is indeed the case for effective quantum gravity.

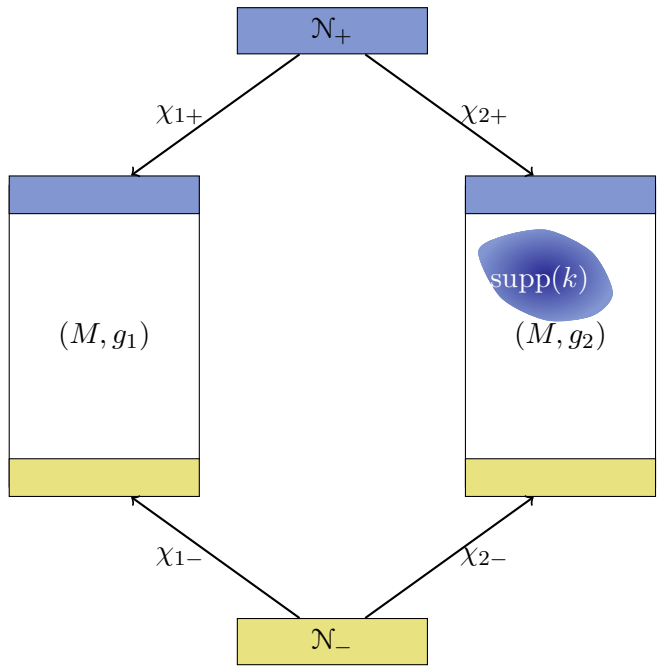
\begin{figure}[h!t]
	\begin{center}
		\begin{tikzpicture}[scale=1.5]
		\draw (-1,0) node[name=m1,shape=rectangle,minimum height=4cm,minimum width=2.5cm,draw] {$(M,g_1)$};
		\draw (-1,1.25) node[name=up1,fill=blue1,shape=rectangle,minimum height=0.5cm,minimum width=2.5cm,draw] {};
		\draw (-1,-1.25) node[name=down1,fill=honey,shape=rectangle,minimum height=0.5cm,minimum width=2.5cm,draw] {};
		\draw (3,0) node[name=m2,shape=rectangle,minimum height=4cm,minimum width=2.5cm,draw] {$(M,g_2)$};
		\draw (3,1.25) node[name=up2,fill=blue1,shape=rectangle,minimum height=0.5cm,minimum width=2.5cm,draw] {};
		\draw (3,-1.25) node[name=down2,fill=honey,shape=rectangle,minimum height=0.5cm,minimum width=2.5cm,draw] {};
		\draw (1,2.7) node[name=N1,fill=blue1,shape=rectangle,minimum height=0.5cm,minimum width=2.5cm,draw] {$\Ncal_+$};
		\draw (1,-2.7) node[name=N2,fill=honey,shape=rectangle,minimum height=0.5cm,minimum width=2.5cm,draw] {$\Ncal_-$};
		\shade[shading=radial, inner color=darksee,outer color=blue1,xshift=25,scale=0.2] plot[smooth cycle] coordinates{(10,0.7)(8,1.7)(7,3.7)(8,4.4)(10,4.7)(12,3.7)(12.9,1.9)(12,0.9)};
		\draw(2,.5) node [xshift=35,color=white]{$\supp (k)$};
		\path[<-] (up1) edge [thick] node[left] {$\chi_{1+}$} (N1);
		\path[<-] (up2) edge [thick] node[right] {$\chi_{2+}$} (N1);
		\path[<-] (down1) edge [thick] node[left] {$\chi_{1-}$} (N2);
		\path[<-] (down2) edge [thick] node[right] {$\chi_{2-}$} (N2);
		\end{tikzpicture}
		\caption{Embeddings of neighborhoods of Cauchy surfaces into spacetimes $\Mcal_1=(M,g_1)$ and $\Mcal_2=(M,g_2)$.\label{RelCE}}
	\end{center}
\end{figure}

In Ref.~\cite{BFgrav} it was conjectured that a condition of background independence can be formulated by means of relative Cauchy evolution. Let us briefly explain what that means. We fix a globally hyperbolic spacetime $\Mcal_1=(M,g_1)$ and choose $\Sigma_-$ and $\Sigma_+$, two Cauchy surfaces in  $\Mcal_1$, such that $\Sigma_+$ is in the future of $\Sigma_-$. Take another globally hyperbolic metric $g_2$ on $M$, such that $k\doteq g_2-g_1$ is compactly supported and its support $K$ lies between  $\Sigma_-$ and $\Sigma_+$.  Now, take two globally hyperbolic spacetimes $\Ncal_{\pm}$ that embed into $\Mcal_1$ and $\Mcal_2$, via $\chi_{1\pm}$, $\chi_{2\pm}$ in such a way that $\chi_{i\pm}(\Ncal_{\pm})$ are causally convex 
neighborhoods of  $\Sigma_\pm$ in  $\Mcal_i$,  $i=1,2$. This is illustrated on figure \ref{RelCE}. 
We use the time-slice axiom to define isomorphisms $\al_{\chi_{i\pm}}$ and the free relative Cauchy evolution is an automorphism of $\fA(\Mcal_1)$ given by $\beta_{0k}=\al^{\phantom{-1}}_{0\chi_{1-}}\circ\al^{-1}_{0\chi_{2-}}\circ\al^{\phantom{-1}}_{0\chi_{2+}}\circ\al_{0\chi_{1+}}^{-1}$. It was shown in Ref.~\cite{BFV} (see also Ref.~\cite{BFRej}) that the functional derivative of $\beta_{0k}$ with respect to $k$ is equal to the commutator with the free stress-energy tensor. More precisely
\[
\frac{\delta}{\delta k_{\mu\nu}} \beta_{0k}\left(\Scal(F)\right)\Big|_{k=0}=-\frac{i}{\hbar}\left[T_0^{\mu\nu},\Scal(F)\right]_\star\,,
\]
where $T_0^{\mu\nu}$ is the stress-energy tensor of the linearized theory.

To obtain the relative Cauchy evolution for the full interacting theory, we use the Bogoliubov formula \eqref{Bogform} and the free Cauchy evolution $\beta_{0k}$. The resulting automorphism is denoted by $\beta_{k}$ and its functional derivative by $\Theta^{\mu\nu}$. It was shown in Ref.~\cite{BFRej} that
	\[
	(\Theta^{\mu\nu}(F))_{\mathrm{int}}\stackrel{o.s.}{=}-\frac{i}{\hbar}[T^{\mu\nu}_{\mathrm{int}},F_{\mathrm{int}}]_\star\,,
	\]
	where $T^{\mu\nu}_{\mathrm{int}}$ is the interacting stress-energy tensor of the full extended Lagrangian $L$. One can use the renormalization freedom to ensure that
	$T^{\mu\nu}_{\mathrm{int}}=0$ holds, so the interacting theory is background independent.
\section{Conclusions and Outlook}
Recent developments in QFT on curved spacetimes have shown that the algebraic approach to quantum field theory has several advantages that allow to overcome technical and conceptual problems. It also allows us to push the limits of our understanding of the nature of space and time even further and to learn something about the structure of quantum gravity. In particular, in Ref.~\cite{BFRej} it was shown how to construct diffeomorphism invariant observables for gravity which have some local as well as global features and allow for quantization using the Epstein-Glaser renormalization scheme. In the future investigation on can now look at concrete models and see how the QG corrections can be implemented. A natural directions would be cosmology and black hole physics.


\begin{thebibliography}{00}
	\bibitem{BRZ} D. Bahns, K. Rejzner, J. Zahn, {\em Commun. Math. Phys.} {\bf 327}, 779-814 (2014).
	\bibitem{Berg} P. G. Bergmann, {\em Reviews of Modern Physics} {\bf 33}, 510 (1961).
	\bibitem{BergKom} P. G. Bergmann and A. B. Komar, {\em PRL} {\bf 4}, 432 (1960).
	\bibitem{BrK} J. D. Brown  and K. V. Kucha\v{r}, {\em Phys. Rev. D} {\bf 51} 5600 (1995). 
	\bibitem{BDF} R. Brunetti, M. D{\"u}tsch and K. Fredenhagen, {\em Adv. Theor. Math. Phys.} {\bf 13}, 1541-1599 (2009).
	\bibitem{BF0} R. Brunetti, K. Fredenhagen, {\em Commun. Math. Phys.} {\bf 208}, 623-661 (2000).
\bibitem{BFgrav}  R. Brunetti and K. Fredenhagen, in {\em Quantum Gravity, mathematical models and experimental bounds}, B. Fauser, J. Tolksdorf and E. Zeidler, Eds.,  (Springer, 2006) 151-159.
\bibitem{BFRej} {\tt arXiv:gr-qc/9910001}, R. Brunetti,  K. Fredenhagen and K. Rejzner, {\em Quantum gravity from the point of view of locally covariant quantum field theory}, 2013.
\bibitem{BFV} R. Brunetti, K. Fredenhagen and R. Verch, {\em Commun. Math. Phys.} {\bf 237}, 31-68 (2003).
\bibitem{CHP09} A. Coley, S. Hervik and N. Pelavas, {\em Classical and Quantum Gravity} {\bf 26}, 025013 (2009).
\bibitem{DMP09} C. Dappiaggi, V. Moretti and N. Pinamonti, {\em JMP} {\bf 50}, 062304 (2009).
\bibitem{DMP11} C. Dappiaggi, V. Moretti and N. Pinamonti, {\em Adv. Theor. Math. Phys.} {\bf 15}, 355-447 (2011).
\bibitem{Dittrich:2005kc} B. Dittrich, {\em Classical and Quantum Gravity} {\bf 23}, 6155 (2006).
\bibitem{DF}  M. D{\"u}tsch and K. Fredenhagen, in {\em Mathematical Physics in Mathematics and Physics: Quantum and Operator Algebraic Aspect} (AMS, 2001).
\bibitem{EG} H. Epstein and V. Glaser, {\em AHP} {\bf 19}, 211-295 (1973).
\bibitem{FewVerchReview} C. J. Fewster and R. Verch, in {\em Advances in Algebraic Quantum Field Theory} (Springer, 2015), 125-189.
\bibitem{FR}  K. Fredenhagen and K. Rejzner, {\em Commun. Math. Phys.} {\bf 314}, 93-127 (2012).
\bibitem{FRindep}  K. Fredenhagen and K. Rejzner, in {\em Quantum Field Theory and Gravity} (Springer, 2012), 15-23.
\bibitem{FR2}  K. Fredenhagen and K. Rejzner, {\em Commun. Math. Phys.} {\bf 317}, 697-725 (2012).
\bibitem{FR14}  K. Fredenhagen and K. Rejzner, {\em JMP} {\bf 57}, 031101 (2016).
\bibitem{LesHouches}  K. Fredenhagen and K. Rejzner, in {\em Mathematical Aspects of Quantum Field Theories} (Springer, 2015), 17-55.
\bibitem{Weinberg} J. Gomis and S. Weinberg, {\em Nucl. Phys. B} {\bf 469}, 473-487 (1996). 
\bibitem{Haag} R. Haag, {\em Local quantum physics}, 2nd edition (Springer-Verlag, Berlin 1993).
\bibitem{HK} R. Haag and D. Kastler, {\em JMP} {\bf 5}, 848-861 (1964).
\bibitem{Hack14} T.-P. Hack, {\em Classical and Quantum Gravity} {\bf 31}, 215004 (2014).
\bibitem{ThomasBook} T.-P. Hack, {\em Cosmological Applications of Algebraic Quantum Field Theory in Curved Spacetimes} (Springer, 2015).
\bibitem{CH10} S. Hervik and A. Coley, {\em Classical and Quantum Gravity} {\bf 27}, 095014 (2010).
\bibitem{H} S. Hollands, {\em Rev. Math. Phys.} {\bf 20}, 1033-1172 (2008).
\bibitem{HW}  S. Hollands and R. M. Wald, {\em Commun. Math. Phys.} {\bf 223}, 289-326 (2001).
\bibitem{HW01}  S. Hollands and R. M. Wald, {\em Commun. Math. Phys.} {\bf 231}, 309-345 (2002).
\bibitem{Moretti} W. Moretti, {\em Spectral Theory and Quantum Mechanics: With an Introduction to the Algebraic Formulation} (Springer, 2013).
\bibitem{MuSa} O. M{\"u}ller, M. S{\'a}nchez, {\em Transactions of the American Mathematical Society} {\bf 363}, 5367-5379 (2011).
\bibitem{Pei} R. E. Peierls, {\em Proceedings of the Royal Society of London. Series A. Mathematical and Physical Sciences} {\bf 214}, 143-157 (1952). 
\bibitem{Reu98} M. Reuter, {\em Phys. Rev. D} {\bf 57}, 971 (1998).
\bibitem{Reu02} M. Reuter and F. Saueressig, {\em Phys. Rev. D} {\bf 65}, 065016 (2002).
	\bibitem{Rovelli:2001bz} C. Rovelli, {\em Phys. Rev. D} {\bf 65}, 124013 (2002).
	\bibitem{WernerSummers} S. J. Summers and R. Werner, {\em Commun. Math. Phys.} {\bf 110}, 247-259 (1987).
	\bibitem{Thiemann:2004wk} T. Thiemann, {\em Classical and Quantum Gravity} {\bf 23}, 1163 (2006).
\bibitem{Weinberg79} S. Weinberg, in {\em General relativity} (Cambridge University Press, 1979).
\end{thebibliography}
\end{document}